\documentclass[11pt,a4paper,english,showpacs,superscriptaddress,prd,aps,preprint]{revtex4-1}
\usepackage{amssymb}
\usepackage{amsmath} 
\usepackage{graphicx}
\usepackage{babel}
\usepackage{hyperref}
\usepackage{color}

\begin{document}

\title{Effective field theory of quantum gravity coupled to scalar electrodynamics}

\author{L.~Ibiapina~Bevilaqua}
\email{leandro@umass.edu}
\affiliation{Escola de Ci\^encias e Tecnologia, Universidade Federal do Rio Grande do Norte\\
Caixa Postal 1524, 59072-970, Natal, Rio Grande do Norte, Brazil}
\affiliation{Department of Physics, University of Massachusetts, Amherst, Massachusetts 01003, USA}

\author{A.~C.~Lehum}
\email{lehum@ufpa.br}
\affiliation{Faculdade de F\'isica, Universidade Federal do Par\'a, 66075-110, Bel\'em, Par\'a, Brazil.}

\author{A. J. da Silva}
\email{ajsilva@fma.if.usp.br}
\affiliation{Instituto de F\'\i sica, Universidade de S\~ao Paulo\\
Caixa Postal 66318, 05315-970, S\~ao Paulo, S\~ao Paulo, Brazil}

\begin{abstract}
In this work we use the framework of effective field theory to couple Einstein's gravity to scalar electrodynamics and determine the renormalization of the model through the study of physical processes below Planck scale, a realm where quantum mechanics and general relativity are perfectly compatible. We consider the effective field theory up to dimension six operators, corresponding to processes involving one graviton exchange. Studying the renormalization group functions we see that the beta function of the electric charge is positive and possesses no contribution coming from gravitational interaction. Our result indicates that gravitational corrections do not alter the running behavior of the gauge coupling constants, even if massive particles are present.
\end{abstract}



\maketitle

\section{Introduction}

Quantum gravity based in Einstein's theory of gravitation have been the subject of several papers over the last fifty years, including the seminal papers by Feynman \cite{Feynman:1963} and DeWitt \cite{DeWitt:1967}. The quantum aspects of gravity coupled to scalar electrodynamics has also being discussed, for example, in \cite{Odintosov:1990a,Odintosov:1990b}, where the effective action was computed (for a more comprehensive text, see \cite{Odintosov:1992}).

The theory is notoriously nonrenormalizable~\cite{'tHooft:1974bx,PhysRevLett.32.245,Deser:1974cy}, since it requires a set of infinite parameters to absorb all the divergences coming from the loop diagrams. The potential harm of a nonrenormalizability, however, can be overcome in the effective field theory (EFT) framework, where there is a unambiguous way to define a well behaved and reliable quantum theory of general relativity, if only we agree to restrict ourselves to low energies compared to the Planck scale~\cite{Donoghue:1994dn,Burgess:2003}. In the core of EFT argument is the fact that if we respect all the symmetry of the problem to write down in the Lagrangian all the terms that contribute to a process of energy scale $E$, all new terms eventually required by the renormalization procedure can be neglected because they must contribute only to higher energy processes. This EFT's methods has been used to compute several gravitational corrections (see~\cite{Donoghue:2012zc} and references therein).

Motivated by this modern view of quantum gravity, Robinson and Wilczek~\cite{Robinson:2005fj} considered a non-Abelian gauge field coupled to gravity and found the gravity contributes with a negative term to the beta function of the gauge coupling, meaning that quantum gravity could make gauge theories asymptotically free. The origin of this effect would be the arising of quadratic UV divergences, associated to the one graviton exchange graph, that could be absorbed in a gauge coupling constant redefinition. This remarkable conclusion motivated a lot of research on the subject. A few months after Robinson-Wilczek's paper, their conclusion was questioned by Pietrykowski, who repeated their calculation for an abelian field and reproduced their result {\it for a particular gauge choice}, but showed that a different gauge could lead to no gravitational contribution at all~\cite{Pietrykowski:2006xy}. Pietrykowski also suggest at the end of Ref.~\cite{Pietrykowski:2006xy} that if dimensional regularization is applied, the quadratic divergence would not be present, and that claim was investigated further by Felipe {\it et al} in Refs.~\cite{Felipe:2012vq}, where it is argued that the gravitational correction to the beta function computed in~\cite{Robinson:2005fj} is regularization dependent and therefore ambiguous. The absence of gravitational correction at one-loop order was reinforced by Ebert, Plefka and Rodigast in a paper where they follow diagrammatic approach using both a cut-off and a dimensional regularization~\cite{Ebert:2007}. A detailed study of the use of the Vilkovisky-DeWitt method to this problem was done in Ref.~\cite{Nielsen:2012fm}, where Nielsen shows for the Einstein-Maxwell system that quadratic divergences would break the Ward identities, so the method would guarantee only the gauge invariance of finite and logarithmic divergent parts of the effective action.

The role of the cosmological constant was also investigated and the result was that it should induce an asymptotic freedom behavior to the electric charge~\cite{Toms:2008dq,Toms:2010vy,Toms:2011zza} and for $\lambda\phi^4$ model~\cite{Lehum:2013oja}. There has been also some results for other (nongauge) interactions, for example, it was argued that massive particles with Yukawa~\cite{Rodigast:2009zj} and $\phi^4$~\cite{Pietrykowski:2012nc} share the same property of asymptotic freedom, an effect that vanishes when the masses are withdrawn.  

The controversy about the beta function calculations is still unresolved, and some works question the physical meaning of the definition of the running coupling constants~\cite{Donoghue:2012zc,Ellis:2010,Anber:2010uj}, and these papers argue that a scattering matrix computation is needed to give a physical definition for the running of the coupling constants. Using S-matrix, it was found in Ref. ~\cite{Anber:2010uj} that an attempt to compute the running of the Yukawa coupling can be ambiguous, since it would seem to run in the direction of asymptotic freedom in one process, but will increase with energy for another process, and therefore what appears to be asymptotic freedom is not an universal behaviour within the theory. This conclusion is intrinsically related to the construction and meaning of an effective theory: the gravitational corrections will not renormalize the original operator, but rather a higher derivative one (because of the dimensional coupling constant) and different processes typically involve different combinations of operators and no universality is to be expected~\cite{Donoghue:2012zc}. Through the computation of scattering processes, it was shown that quantum gravitational corrections do not alter the running behavior of the electric charge in the massless Scalar QED~\cite{Charneski:2013zja}, but the presence of a positive cosmological constant in the massless Einstein-$\lambda\phi^4$ model corroborates earlier proposal\cite{Toms:2008dq}. 

In this work we investigate the renormalization of the effective field theory of the massive Scalar Electrodynamics coupled to Einstein's Gravity, showing that gravitational corrections do  not alter the running behavior of the electric charge, even in the presence of massive particles. This is done by the computation of scattering amplitudes between charged particles.

\section{Massive Scalar QED coupled to Einstein's Gravity}\label{sec1}

The model is given by the following action
\begin{eqnarray}\label{eq01}
S=\int{d^4x }\sqrt{-g}&&\Big{\{}\frac{2}{\kappa^2}R-\frac{1}{4}g^{\mu\alpha}g^{\nu\beta}F_{\alpha\beta}F_{\mu\nu}+g^{\mu\nu}\left( \partial_\mu +ieA_\mu\right) \phi_j \left(\partial_\nu-ieA_\nu\right)\phi^*_j\nonumber\\
&&-m^2(\phi^*_j\phi_j)-\frac{\lambda}{2}(\phi^*_j\phi_j)^2 +\mathcal{L}_{HO} +\mathcal{L}_{GF}+\mathcal{L}_{CT} \Big{\}},
\end{eqnarray}

\noindent where $j$ assume values $a$ and $b$ according to the flavor of the pion, $\kappa^2=32\pi G=32\pi/M_P^2$, with $M_P$ being the Planck mass and $G$ the Newtonian gravitational constant, $e$ is the electric charge and $\lambda$ a self interaction constant. $\mathcal{L}_{GF}$ is the gauge-fixing plus Faddeev-Popov ghost Lagrangian (for the graviton and the photon) and $\mathcal{L}_{CT}$ is the Lagrangian of counterterms. {Finally}, $\mathcal{L}_{HO}$ is the Lagrangian of higher derivatives terms given by
\begin{eqnarray}\label{eqho}
\mathcal{L}_{HO} &=& \lambda_1 \partial^\mu(\phi_a\phi^*_a)\partial_{\mu}(\phi_b\phi^*_b)+\lambda_2 (\phi_a \partial^\mu \phi^*_a-\partial^\mu\phi_a \phi^*_a)(\phi_b \partial_\mu \phi^*_b-\partial_\mu\phi_b \phi^*_b)\nonumber\\
&&+ \lambda_3\left(\Box\phi_a\phi^*_a\phi_b\phi^*_b +\phi_a\Box\phi^*_a\phi_b\phi^*_b + \phi_a\phi^*_a\Box\phi_b\phi^*_b + \phi_a\phi^*_a\phi_b\Box\phi^*_b\right)+(\cdots),
\end{eqnarray}
\noindent where $(\cdots)$ stands for omitted higher order terms, which are not important to our analysis in this paper.

We must keep in mind that for renormalized Lagrangian we have redefined $\phi_{0j}=Z_{\phi}^{1/2}\phi_j=\sqrt{1+\delta_\phi}\phi_j$ and $A_{0\mu}=Z_A^{1/2}A_{\mu}=\sqrt{1+\delta_A}A_{\mu}$. The relation between bare and renormalized {coupling constants} are given by
\begin{eqnarray}
e_0&=&\mu^{\epsilon}\frac{Z_e}{Z_\phi Z_{A}^{1/2}}e=\frac{(e+\delta_e)}{Z_\phi Z_{A}^{1/2}}\mu^{\epsilon},\nonumber\\
\lambda_0&=&\mu^{2\epsilon}\frac{Z_\lambda}{Z_\phi^2}\lambda=\frac{(\lambda+\delta_\lambda)}{Z_\phi^2}\mu^{2\epsilon},
\end{eqnarray}

\noindent where $\mu$ is a mass scale introduced by the dimensional regularization with $D=4-2\epsilon$.

Let us consider small fluctuations around the flat metric, i.e.,  
\begin{eqnarray}\label{eq02}
&&g_{\mu\nu}=\eta_{\mu\nu}+\kappa h_{\mu\nu},\\
&&g^{\mu\nu}=\eta^{\mu\nu}-\kappa h^{\mu\nu}+\kappa^2 h^{\mu\alpha}{h_{\alpha}}^{\nu}+\mathcal{O}(\kappa^3),\\
&&\sqrt{-g}=1+\frac{1}{2}\kappa h-\frac{1}{4}\kappa^2h_{\alpha\beta}P^{\alpha\beta\mu\nu}h_{\mu\nu}+\mathcal{O}(\kappa^3),
\end{eqnarray}
\noindent where $\eta_{\mu\nu}=(+,-,-,-)$, $P^{\alpha\beta\mu\nu}=\dfrac{1}{2}(\eta^{\alpha\mu}\eta^{\beta\nu}+\eta^{\alpha\nu}\eta^{\beta\mu}-\eta^{\alpha\beta}\eta^{\mu\nu})$ and $h=\eta^{\mu\nu}h_{\mu\nu}$. For more details see for instance~\cite{Choi:1994ax}.

Through the harmonic gauge-fixing function, $G_\mu=\partial^\nu h_{\mu\nu}-\dfrac{1}{2}\partial_\mu h$, the graviton propagator can be cast as
\begin{eqnarray}\label{eq04}
\langle T~h^{\alpha\beta}(p) h^{\mu\nu}(-p)\rangle&=&D^{\alpha\beta\mu\nu}(p)=\frac{i}{p^2}\left[P^{\alpha\beta\mu\nu}+(\xi_h-1)\frac{Q^{\alpha\beta\mu\nu}(p)}{p^2}
\right],
\end{eqnarray}
\noindent where $Q^{\alpha\beta\mu\nu}(p)=(\eta^{\alpha\mu}p^{\beta}p^\nu+\eta^{\alpha\nu}p^{\beta}p^\mu+\eta^{\beta\mu}p^{\alpha}p^\nu+\eta^{\beta\nu}p^{\alpha}p^\mu)$.

The propagators for the other fields are also obtained by the usual Faddeev-Popov method, resulting in
\begin{eqnarray}\label{eq05}
\langle T~A^\mu(p) A^\nu(-p)\rangle&=&\Delta^{\mu\nu}(p)=-\frac{i}{p^2}\left[ \eta^{\mu\nu}+(1-\xi_\gamma)\frac{p^\mu p^\nu}{p^2}\right],\\
\langle T~\phi^*_i(p) \phi_j(-p)\rangle &=&\Delta_{ij}(p)=\frac{i}{p^2-m^2}\delta_{ij}.
\end{eqnarray}

\noindent The ghost propagators are not useful in the order we are working.The above propagators were written for generic gauge parameters: $\xi_h$ and $\xi_\gamma$. Since our interest is in the study of gauge independent quantities, from now on, we will restrict to the Feynman gauges, $\xi_h=\xi_\gamma=1$, which simplifies the very long calculations involved. Reproduction of the same results by calculating with generic parameters $\xi_h$ and $\xi_\gamma$ would be desirable, but we will leave it for future investigations.

\section{Scattering amplitudes and the running of the coupling constants} \label{sec3}

\begin{figure}[ht]
	\includegraphics[angle=0 ,width=14cm]{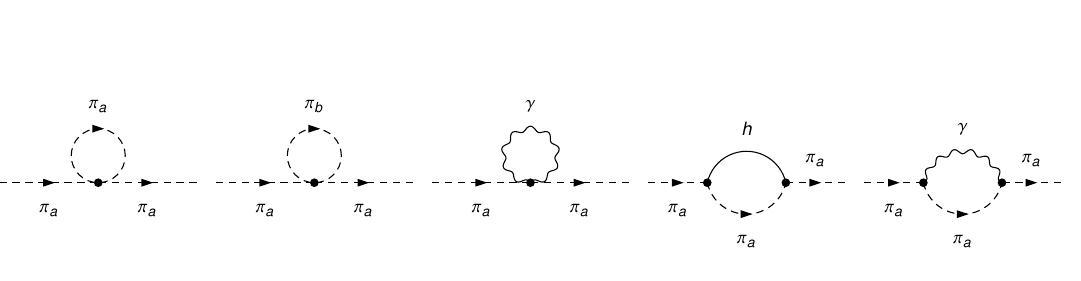}
	\caption{Feynman diagrams for the pions self-energy. Dashed, wavy and continuous lines represent the scalar, photon and graviton propagators, respectively.}
	\label{se1}
\end{figure}

The self-energy process of the scalar particle, Figure \ref{se1}, is
\begin{eqnarray}\label{se01} 
\tau_{2}&=&(1+\delta_\phi)(p^2-m^2-\delta_{m^2})-\Sigma_1(p,m,\lambda,\kappa,e),
\end{eqnarray} 
\noindent where $\Sigma_1(p,m,\lambda,\kappa,e)$ is the one-loop correction given by 
\begin{eqnarray}\label{se02} 
\Sigma_1=-(p^2-m^2)\left(\frac{e^2}{8\pi^2\epsilon}-\frac{\kappa^2m^2}{16\pi^2\epsilon}\right)+\frac{3(\lambda-e^2) m^2}{16\pi^2\epsilon}.
\end{eqnarray}
	
Using the MS renormalization scheme~\cite{minsub,minsub1}, the counterterms for mass ($\delta_{m^2}$) and wave-function ($\delta_\phi$) are given by
\begin{eqnarray}
\delta_{m^2}&=&\frac{3(e^2-\lambda)m^2}{16\pi^2\epsilon},\label{ctm}\\
\delta_\phi&=&\frac{e^2}{8\pi^2\epsilon}-\frac{m^2\kappa^2}{16\pi^2\epsilon}.\label{ctz} 
\end{eqnarray}

\begin{figure}[ht]
	\includegraphics[angle=0 ,width=14cm]{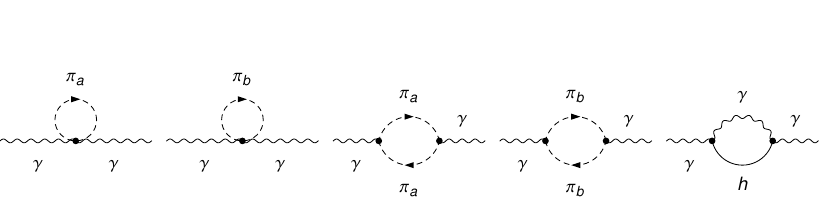}
	\caption{Feynman diagrams for the photon self-energy. Dashed, wavy and continuous lines represent the scalar, photon and graviton propagators, respectively.}
	\label{se2}
\end{figure}

The one-loop correction to the polarization tensor of the photon field, Figure \ref{se2}, is
\begin{eqnarray}\label{se02} 
\Pi^{\mu\nu}=-\frac{e^2}{24 \pi^2 \epsilon }(\eta^{\mu\nu}p^2-p^\mu p^\nu)+H.O.,
\end{eqnarray}

\noindent where $H.O.={\displaystyle -\frac{e^2 \kappa^2}{96 \pi^2 \epsilon }(\eta^{\mu\nu}p^2-p^\mu p^\nu) p^2}$, see for instance~\cite{Ebert:2007}, is a high order gravitational correction which only contributes to the renormalization of a high order operator. From the above expression we find the photon wave-function counterterm as
\begin{eqnarray}\label{se03}
\delta_{A}&=&-\frac{e^2}{24\pi^2\epsilon}.
\end{eqnarray} 

\begin{figure}[ht]
	\includegraphics[height=1cm ,angle=0 ,width=8cm]{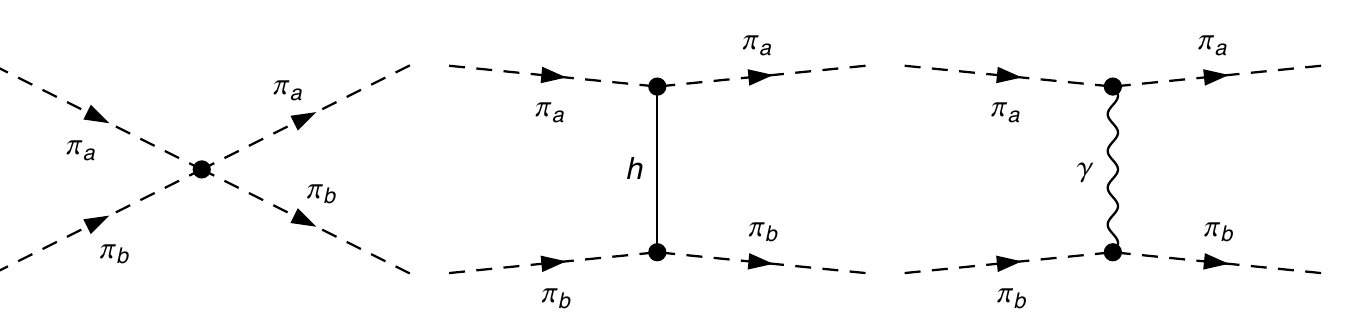}
	\caption{Feynman diagrams for the pion scattering amplitude ($\pi_a^+ +\pi_b^+\rightarrow \pi_a^+ +\pi_b^+$) at tree level.}
	\label{scat01}
	\end{figure}

\begin{figure}[ht]
	\includegraphics[height=1cm ,angle=0 ,width=16cm]{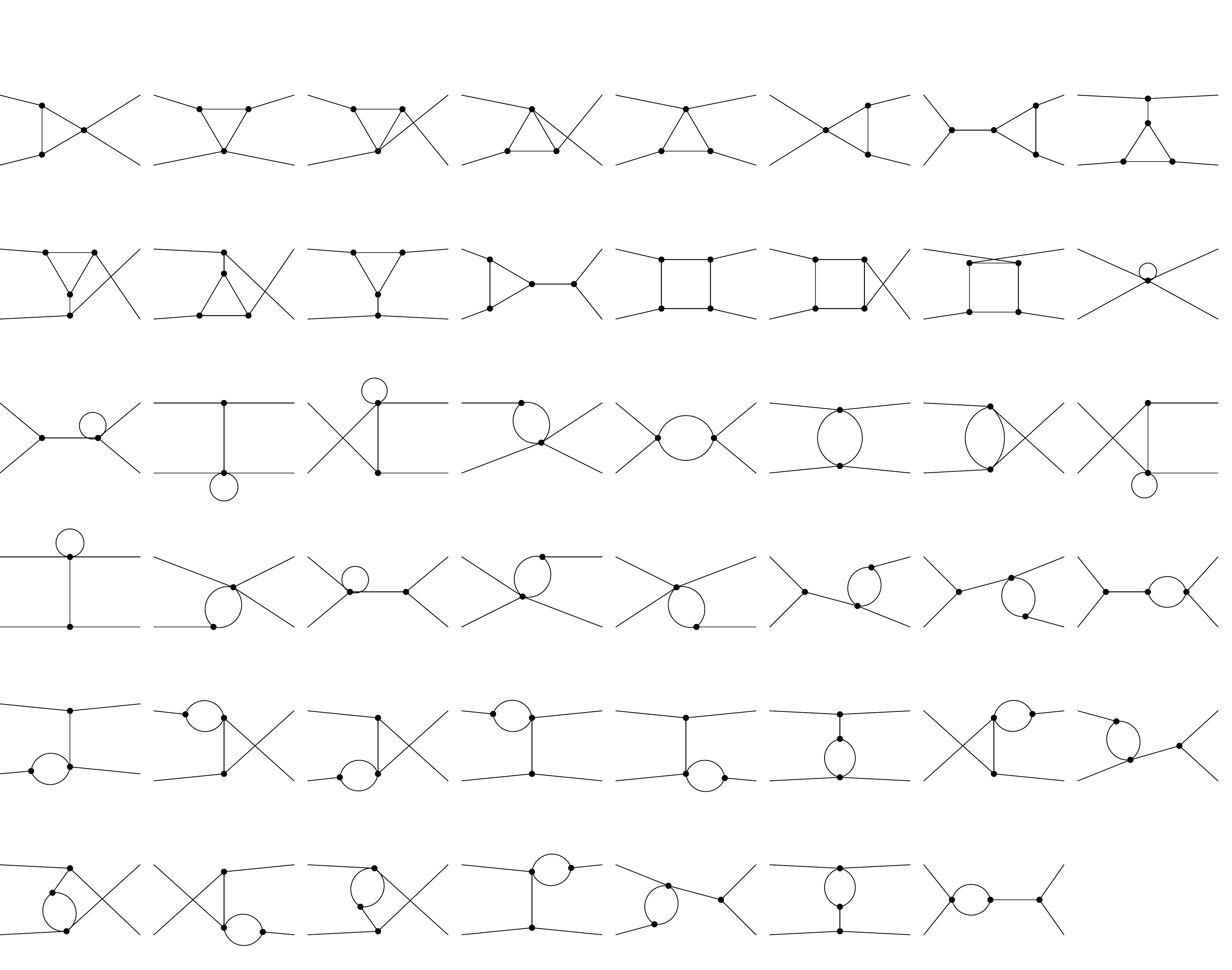}
	\caption{Topologies to the ($\pi_a^+ +\pi_b^+\rightarrow \pi_a^+ +\pi_b^+$) scattering amplitude at one-loop order. After to consider the vertices, we have more than 80 Feynman diagrams to this reaction up to ${\mathcal{O}}(\kappa^2)$.}
	\label{scat02}
	\end{figure}

The scattering amplitude $\pi^+_a + \pi^+_b\rightarrow \pi^+_a + \pi^+_b$, Figures \ref{scat01} and \ref{scat02}, is given by
\begin{eqnarray}\label{amp01}
\mathcal{M}&=&\mathcal{M}_{tree}+\mathcal{M}_{CT}+\mathcal{M}_{1l}\nonumber\\
&=&-\lambda+\frac{e^2 (S-U)}{T}+\frac{\kappa^2}{4}\frac{S U}{T}+\frac{m^2 \kappa^2}{2}-\frac{m^4 \kappa^2}{2 T}+\mathcal{M}_{CT}+\mathcal{M}_{1l},
\end{eqnarray}

\noindent For convenience, we have set $\lambda_1=\lambda_2=\lambda_3=0$ at tree level. In the above expression, $U$, $T$ and $S$ are the Mandelstam variables, $\mathcal{M}_{CT}$ is the expression for the counterterms 
\begin{eqnarray}\label{eqtree01b}
\mathcal{M}_{CT}&=&-\delta_\lambda+\frac{2e\delta_e (S-U)}{T}+\frac{\delta_{\kappa^2}}{4}\frac{S U}{T}+\frac{\delta_{m^2}\kappa^2}{2}+\frac{m^2}{2} \delta_{\kappa^2}\nonumber\\
&&-\frac{ m^2\delta_{m^2}\kappa^2}{T} -\frac{m^4\delta_{\kappa^2}}{2T}
+\delta_{\lambda_1} T+\delta_{\lambda_2} (S-U)-4m^2\delta_{\lambda_3},
\end{eqnarray}

\noindent and $\mathcal{M}_{1l}$ the one-loop correction. To compute $\mathcal{M}_{1l}$ up to order of $\kappa^2$ (one-graviton exchange), we used a set of Mathematica$^{\copyright}$ packages~\cite{feynarts,feyncalc,formcalc} and found:
\begin{eqnarray}\label{eqloop01}
\mathcal{M}_{1l}&=& \frac{3 \lambda ^2}{8 \pi^2 \epsilon }+\frac{3 e^4}{8 \pi^2 \epsilon }-\frac{e^4 S}{4 \pi^2 T \epsilon }+\frac{e^4 U}{4 \pi^2 T \epsilon }-\frac{11 e^2 S \kappa^2}{384 \pi^2 \epsilon }+\frac{e^2 m^4 \kappa^2}{48 \pi^2 T \epsilon }+\frac{7 e^2 m^2 S \kappa^2}{48 \pi^2 T \epsilon }\nonumber\\
&&+\frac{5 e^2 S^2 \kappa^2}{384 \pi^2 T \epsilon }+\frac{11 e^2 T \kappa^2}{192 \pi^2 \epsilon }+\frac{11 e^2 U \kappa^2}{384 \pi^2 \epsilon }+\frac{5 e^2 m^2 U \kappa^2}{24 \pi^2 T \epsilon }-\frac{11 e^2 S U \kappa^2}{96 \pi^2 T \epsilon }-\frac{25 e^2 U^2 \kappa^2}{384 \pi^2 T \epsilon }\nonumber\\
&&-\frac{e^2 \lambda }{8 \pi^2 \epsilon }-\frac{m^2 \kappa^2 \lambda }{8 \pi^2 \epsilon }-\frac{S \kappa^2 \lambda }{32 \pi^2 \epsilon }-\frac{3 m^4 \kappa^2 \lambda }{16 \pi^2 T \epsilon }-\frac{3 T \kappa^2 \lambda }{64 \pi^2 \epsilon }-\frac{U \kappa^2 \lambda }{32 \pi^2 \epsilon }+\mathrm{finite~terms}
\end{eqnarray}

We use also Eq. (\ref{ctm}) to write the amplitude (\ref{amp01}) as
\begin{eqnarray}\label{amp02}
\mathcal{M}&=&-\lambda+\frac{e^2 (S-U)}{T}+\frac{\kappa^2}{4}\frac{S U}{T}+\frac{m^2 \kappa^2}{2}-\frac{m^4 \kappa^2}{2 T}-\delta_\lambda+\frac{2e\delta_e (S-U)}{T}\nonumber\\
&&+\frac{\delta_{\kappa^2}}{2}\frac{S U}{T}+\frac{m^2}{2}\delta_{\kappa^2}-\frac{m^4\delta_{\kappa^2}}{2T}
+\delta_{\lambda_1} T+\delta_{\lambda_2} (S-U)-4m^2\delta_{\lambda_3}\nonumber\\
&&+\frac{(3 \lambda ^2+3e^4-\lambda e^2)}{8 \pi^2 \epsilon }-\frac{e^4 (S-U)}{4 \pi^2 T \epsilon }-\frac{11 e^2 \kappa^2(S-U)}{384 \pi^2 \epsilon }+\frac{(e^2-9\lambda)  m ^4 \kappa^2}{48 \pi^2 T \epsilon }\nonumber\\
&&+\frac{7 e^2  m ^2 \kappa^2 S}{48 \pi^2 T \epsilon }
+\frac{5 e^2 \kappa^2 (S^2-5U^2)}{384 \pi^2 T \epsilon }
+\frac{11 e^2\kappa^2 T}{192 \pi^2 \epsilon }
+\frac{5 e^2  m ^2 \kappa^2 U}{24 \pi^2 T \epsilon }\nonumber\\
&&-\frac{11 e^2 \kappa^2 S U}{96 \pi^2 T \epsilon }-\frac{ m ^2 \kappa^2 \lambda }{8 \pi^2 \epsilon }-\frac{\kappa^2 \lambda (S+U+T) }{32 \pi^2 \epsilon }-\frac{ \kappa^2 \lambda T }{64 \pi^2 \epsilon } +\mathrm{finite~terms}.
\end{eqnarray}

The kinematical identity $S+T+U=4m^2$ can then be used to eliminate $U$ from $\mathcal{M}$ in order to simplify the analysis. Collecting terms with the same kinematical factor, we have
\begin{eqnarray}\label{amp03}
\mathcal{M}&=&-\lambda+e^2+\frac{m^2 \kappa^2}{2}+2 e \delta_e +\frac{m^2 \delta_{\kappa^2} }{2}-\delta_\lambda -4 m^2 \delta_{\lambda_2}-4 m^2 \delta_{\lambda_3}+\frac{e^4}{8 \pi^2 \epsilon }\nonumber\\
&&+\frac{25 e^2 m^2 \kappa^2}{48 \pi^2 \epsilon } -\frac{e^2 \lambda }{8 \pi^2 \epsilon }-\frac{11 m^2 \kappa^2 \lambda }{32 \pi^2 \epsilon }+\frac{3 \lambda ^2}{8 \pi^2 \epsilon }+\left(-\frac{\delta_{\kappa^2} }{4}-\frac{\kappa^2}{4}+\frac{e^2 \kappa^2}{16 \pi^2 \epsilon }\right)\frac{S^2}{T}\nonumber\\
&&+\left( -4 e^2 m^2-8 e m^2 \delta_e -\frac{m^4 \delta_{\kappa^2} }{2}+\frac{2 e^4 m^2}{\pi^2 \epsilon }-\frac{m^4 \kappa^2}{2}-\frac{3 e^2 m^4 \kappa^2}{8 \pi^2 \epsilon }   \right) \frac{1}{T}\nonumber\\
&&+\left(-\frac{\delta_{\kappa^2} }{4}+2 \delta_{\lambda_2}-\frac{\kappa^2}{4}-\frac{7 e^2 \kappa^2}{96 \pi^2 \epsilon }  \right) S
+\left( \delta_{\lambda_1}+\delta_{\lambda_2}-\frac{7 e^2 \kappa^2}{192 \pi^2 \epsilon }-\frac{\kappa^2 \lambda }{64 \pi^2 \epsilon } \right)T \nonumber\\
&&+\left(2 e^2+4 e \delta_e +m^2 \delta_{\kappa^2} -\frac{e^4}{2\pi^2 \epsilon }+m^2 \kappa^2  \right)\frac{S}{T}
+\mathrm{finite~terms}.
\end{eqnarray}

Imposing finiteness, we find
\begin{eqnarray}
	&&\delta_e=\frac{e^3}{8\pi^2\epsilon}-\frac{e\kappa^2 m^2}{16\pi^2\epsilon},\label{ct01}\\
	&&\delta_{\kappa^2}=\frac{\kappa^2e^2}{4\pi^2\epsilon},\label{ct02}\\ 
	&&	\delta_{\lambda_1}=\frac{\kappa^2(\lambda-2e^2)}{64\pi^2\epsilon},\label{ct03}\\
	&&\delta_{\lambda_2}=\frac{13\kappa^2e^2}{192\pi^2\epsilon}.\label{ct04}
	\end{eqnarray}
	
Since $\displaystyle{e_0=\mu^{\epsilon}e\frac{Z_e}{Z_\phi Z_A^{1/2}}=\mu^{\epsilon}\frac{(e+\delta_e)}{(1+\delta_\phi)(1+\delta_A)^{1/2}}}$, and $\displaystyle{\delta_\phi=\frac{\delta_e}{e}=\frac{e^2}{8\pi^2\epsilon}-\frac{\kappa^2 m^2}{16\pi^2\epsilon}}$, we have $\displaystyle{e_0=e\mu^{\epsilon}Z_A^{-1/2}}$. Therefore, the beta function of the electric charge is 
\begin{eqnarray}
\beta(e)=\mu\frac{de}{d\mu}=\frac{e^3}{48 \pi^2},
\end{eqnarray}
\noindent which is simply the usual beta function in absence of gravity, {so we can say that} quantum gravitational corrections do not alter the running behavior of the electric charge.

On the other hand, for the renormalization of $\lambda$, we find
\begin{eqnarray}
&&-\delta_\lambda-4m^2\delta_{\lambda_3}+\frac{(3\lambda^2+3e^4-\lambda e^2)}{8\pi^2\epsilon}+\frac{e^2\kappa^2 m^2}{4\pi^2\epsilon}-\frac{11\lambda\kappa^2m^2}{32\pi^2\epsilon}=0,
\end{eqnarray}
\noindent and we see that we cannot separate the contributions for $\delta_\lambda$ and $\delta_{\lambda_3}$.

This arbitrariness in the definition of the coupling constants is due to the mixing of operators, typical in effective field theories, as discussed in \cite{Anber:2010uj}. The separation of the renormalization of  $\lambda$ and $\lambda_3$ would require the study of off-shell processes where these two parameters, or at least one of them, are involved. For the moment, we will not pursuit this analysis.

\section{Final Remarks}\label{summary}
 
In summary, we have shown that massive scalar QED coupled to gravity is renormalizable within the framework of effective field theory. The appropriate counterterms were set by imposing finiteness of the scattering amplitude at one-loop order.

Our attention was focused on the gravitational correction to the renormalization of the coupling constants, since previous works have indicated that the presence of a dimensionful parameter (the cosmological constant in \cite{Toms:2008dq}) might render a non-vanishing new term. We found that the counterterm for the electric charge does receive a gravitational contribution in the presence of a dimensionful parameter (the mass of the scalars). However, this dependence on $\kappa^2$ is exactly canceled when we compute $\beta(e)$, such that, the running of the electric charge is not altered by gravitational effects, similarly to what was observed in massless QED coupled to gravity\cite{Charneski:2013zja}. 

On the other hand, the counterterms for the scalar coupling constants $\lambda$ and $\lambda_3$ will depend on the gravitational coupling $\kappa^2$ in the case considered ($m\neq0$) and cannot be separated, a manifestation of what is called mixing of operators (typical of effective field theories, as discussed in~\cite{Anber:2010uj}). Their separation would requires the study of off-shell processes involving these coupling constants. For the moment, we will not pursuit this analysis.

\acknowledgments
This work was partially supported by Funda\c{c}\~ao de Amparo \`a Pesquisa do Estado de S\~ao Paulo (FAPESP), Conselho Nacional de Desenvolvimento Cient\'{\i}fico e Tecnol\'{o}gico (CNPq) and Funda\c{c}\~{a}o de Apoio \`{a} Pesquisa do Rio Grande do Norte (FAPERN).

\vspace{2cm}

\end{document}